\documentclass[a4paper,twocolumn,superscriptaddress,11pt,accepted=2020-09-18]{quantumarticle}
\pdfoutput=1

\usepackage{epsfig}
\usepackage{amsfonts}
\usepackage{amsmath}
\usepackage{amssymb}
\usepackage{amsthm}
\usepackage{color}
\usepackage{multirow}
\usepackage[normalem]{ulem}
\newcommand{\stkout}[1]{\ifmmode\text{\sout{\ensuremath{#1}}}\else\sout{#1}\fi}
\usepackage{latexsym}
\usepackage{mathrsfs}
\usepackage[numbers,sort&compress]{natbib}
\usepackage{verbatim}
\usepackage[T1]{fontenc}
\usepackage{float}

\usepackage{graphicx}
\usepackage{xcolor}

\usepackage[colorlinks=true,linkcolor=blue,citecolor=magenta,urlcolor=blue]{hyperref}

\DeclareMathOperator{\Tr}{tr}

\newcommand{\ket}[1]{|#1\rangle}

\begin{document}
	
	
\title{Informationally restricted quantum correlations}

	
\author{Armin Tavakoli}
\affiliation{D\'epartement de Physique Appliqu\'ee, Universit\'e de Gen\`eve, CH-1211 Gen\`eve, Switzerland}

\author{Emmanuel Zambrini Cruzeiro}
\affiliation{D\'epartement de Physique Appliqu\'ee, Universit\'e de Gen\`eve, CH-1211 Gen\`eve, Switzerland}

\author{Jonatan Bohr Brask}
\affiliation{Department of Physics, Technical University of Denmark, Fysikvej, 2800 Kongens Lyngby, Denmark}

\author{Nicolas Gisin}
\affiliation{D\'epartement de  Physique Appliqu\'ee, Universit\'e de Gen\`eve, CH-1211 Gen\`eve, Switzerland}

\author{Nicolas Brunner}
\affiliation{D\'epartement de  Physique Appliqu\'ee, Universit\'e de Gen\`eve, CH-1211 Gen\`eve, Switzerland}

\begin{abstract}
Quantum communication leads to strong correlations, that can outperform classical ones. Complementary to previous works in this area, we investigate correlations in prepare-and-measure scenarios assuming a bound on the information content of the quantum communication, rather than on its Hilbert-space dimension. Specifically, we explore the extent of classical and quantum correlations given an upper bound on the one-shot accessible information. We provide a  characterisation of the set of classical correlations and show that quantum correlations are stronger than classical ones. We also show that limiting information rather than dimension leads to stronger quantum correlations. Moreover, we present device-independent tests for placing lower bounds on the information given observed correlations. Finally, we show that  quantum communication carrying $\log d$ bits of information is at least as strong a resource as $d$-dimensional classical communication assisted by pre-shared entanglement.  
\end{abstract}
	
		
\maketitle
	
	
\section{Introduction} 

Separated parties, initially independent, can become correlated via communication. Intuitively, more communication enables stronger correlations. Also, the strength of the correlations may vary depending on the nature of the communication; for example if the message is carried by a quantum system rather than a classical one. In general, understanding the relation between communication and correlations is a fundamental question, at the intersection of information theory and physics. 

\begin{figure}
	\includegraphics[width=0.85\columnwidth]{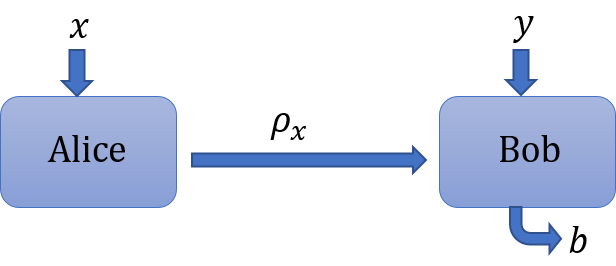}
	\caption{Prepare-and-measure scenario. In this work we investigate the strength of possible correlations $p(b|x,y)$ given a limit on the information carried by the quantum message $\rho_x$.}\label{FigScenario}
\end{figure}

Consider a simple scenario (see Fig.~\ref{FigScenario}) with two separated parties. A first party, Alice, receives an input $x$ and sends a message to a second party, Bob. Upon receiving this message, as well  as some input $y$, Bob produces an output $b$. When repeated many times (with inputs $x$ and $y$ randomly sampled), this experiment is described by the conditional probability distribution $p(b|x,y)$ which characterises the correlations between Alice and Bob. Clearly, the amount of information about $x$ encoded in Alice's message determines the strength of the possible correlations. If Alice sends no message at all (or if the message is independent of $x$), then no correlations are generated, i.e.~$p(b|x,y)=p(b|y)$. On the other hand, if the message perfectly encodes $x$, then maximal correlations can be established; any distribution $p(b|x,y)$ is possible. Thus the main question is: how strong correlations can be established provided that the amount of communication from Alice to Bob is quantitatively limited?

Naturally, the answer depends on how exactly communication is quantified. The most common approach consists in measuring communication via the dimension of the message, i.e.~the number of bits the message could carry. This is used in the field of communication complexity (see e.g. \cite{CCPreview}), where the goal is to find out how the minimum dimension required to solve a problem (i.e.~demanding that the output $b$ corresponds to a certain function of the inputs $x$ and $y$) scales with the problem size. Notably, the use of quantum communication is advantageous since it allows one to solve certain problems with exponentially smaller dimension \cite{Buhrman98, Raz}. In parallel, there has been interest in characterising the set of possible correlations $p(b|x,y)$ for classical and quantum systems of bounded dimension \cite{Gallego, Hendrych, Ahrens, NV}. Again, quantum correlations turn out to be stronger than classical ones. This led to a novel framework for quantum information processing termed ``semi-device-independent'' \cite{Pawlowski, LY11, WP, Lunghi, TK18}, where devices are assumed to process quantum systems of bounded dimension, but are otherwise uncharacterised.

However, measuring communication via the dimension provides only a partial characterisation. Information-theoretic concepts are typically better suited to get a complete picture. This raises a natural question, namely to understand the relation between the strength of correlations and the amount of information that the communication contains. But then, information about what? In correlation experiments, the answer is very natural: we are interested in the information that the message contains about Alice's input $x$.

Here we formalise this problem and investigate classical and quantum correlations for informationally restricted communication. Naturally, however, there are many different ways of quantifying information based on entropies. We quantify the information content of an ensemble of prepared states (classical or quantum) via a one-shot version of accessible information based on min-entropies \cite{Info}. This choice of information measure has a two-fold motivation. Firstly, it admits a simple operational interpretation in terms of how well one could determine Alice's input from her message, via the best possible measurement. Secondly, it proves convenient to work with. Our approach is clearly complementary to previous works based on dimension. Firstly, information is a continuous quantity, while dimension is discrete; one can consider ensembles of states carrying only half a bit of information about Alice's input, which would have no analogue using dimension. Secondly, even when considering ensembles of states carrying $\log d$ bits of information (for some dimension $d$), there exist ensembles of dimension $d'>d$ that carry no more than $\log d$ bits of information, e.g. certain ensembles of non-orthogonal quantum states.

In this work, we develop a framework for characterising informationally restricted correlations. For the case of classical systems, we show that the relevant set of correlations forms a convex polytope, which can be fully characterised. This allows one to find  the minimal amount of information required to reproduce a given correlation using classical communication. In turn, we prove that quantum correlations can be stronger than classical ones. Moreover, we derive device-independent lower bounds on the information, given observed correlations. These ideas are illustrated in a simple scenario. We also show that ensembles of higher-dimensional quantum states carrying no more than one bit of information can generate stronger correlations than two-dimensional quantum systems (i.e.~qubits). Finally, we show that any correlations achievable with classical communication (of a $d$-dimensional message) assisted by pre-shared entanglement can also be achieved using quantum communication carrying $\log d$ bits of information. 

\section{Setting} We start by defining informationally restricted correlations in a quantum prepare-and-measure scenario. The sender, Alice, receives an input $x\in[n]$ sampled from a random variable $X$ (where $[s]=\{1,\ldots,s\}$) which she encodes into a quantum state $\rho_x$ that she relays to the receiver, Bob. Bob also receives a random input $y\in[l]$ and then measures the received state with some generalised measurement (positive operator-valued measure, POVM) $\{M_{b|y}\}$ with outcome $b\in[k]$. The observed correlations are
\begin{equation} 
p(b|x,y)=\Tr\left(\rho_x M_{b|y}\right).
\end{equation}

Let us now characterise the information in Alice's message about her input $x$. Since $x$ is random, sampled from some distribution $p_X(x)$, the ensemble of messages is given by $\mathcal{E}=\{p_X(x),\rho_x\}$. How well could an observer, via any possible POVM $\{N_z\}$, guess $x$ from $\mathcal{E}$? The \textit{guessing probability} is 
\begin{equation}\label{guessing}
P_g(X|\mathcal{E})=\max_{\{N_z\}} \sum_{x=1}^{n} p_X(x)\Tr\left[\rho_xN_x\right].
\end{equation}
Note that the optimal POVM, $\{N_z^*\}$, does not need to be part of set of POVMs $\{M_{b|y}\}$. Hence the statistics obtained from measuring $\{N_z^*\}$ do not necessarily appear in the correlations $p(b|x,y)$.

The observer's minimal uncertainty about $X$ when provided $\mathcal{E}$, i.e.~the conditional min-entropy, is  $H_{\text{min}}(X|\mathcal{E})=-\log\left[P_g(X|\mathcal{E})\right]$. The amount of information gained by learning $\mathcal{E}$, i.e.~the information carried by $\mathcal{E}$,  is then the difference in uncertainty without and with the communication \cite{Info}; 
\begin{equation}\label{comm}
\mathcal{I}_X(\mathcal{E})=H_{\text{min}}(X)-H_{\text{min}}(X|\mathcal{E}),
\end{equation}
where $H_{\text{min}}(X)=-\log\left[\max_x p_X(x)\right]$ is the min-entropy. The quantity $\mathcal{I}_X(\mathcal{E})$ can be viewed as a single-shot version of accessible information \cite{Holevo, JozsaRobbWootters1994}. Note that for any given ensemble $\mathcal{E}$, the guessing probability (and hence the information) can be computed via a semidefinite program \cite{Boyd}. 

We can now define the set of possible correlations $p(b|x,y)$ when the information of the message is upper bounded. Importantly, we do not limit the Hilbert-space dimension for representing the set of the quantum states $\{\rho_x\}$. We also allow for shared randomness between Alice's and Bob's devices. This makes the model more general, and at the same time simplifies the characterisation of the sets of correlations (as these sets are now convex). Formally, we define the set $\mathcal{S}^\text{Q}_\alpha$ of correlations of the form
\begin{equation}
p(b|x,y)=\sum_{\lambda}p(\lambda) \Tr\left(\rho_x^{(\lambda)}M_{b|y}^{(\lambda)}\right),
\end{equation}
where $\lambda$ denotes the shared classical variable, distributed according to $p(\lambda)$, and the information is bounded by $\mathcal{I}_X\leq \alpha$. The quantity $\mathcal{I}_X$ is computed via Eq. \eqref{comm}, considering the average guessing probability of the ensemble $\mathcal{E}=\{p(\lambda),\mathcal{E}_\lambda\}$:
\begin{equation}\label{SR}
P_g(X|\mathcal{E})=\sum_{\lambda}p(\lambda)P_g(X|\mathcal{E}_\lambda),
\end{equation}
where $P_g(X|\mathcal{E}_\lambda)$ denotes the guessing probability for the subensemble $\mathcal{E}_\lambda=\{p_X(x),\rho_x^{(\lambda)}\}$.

\section{Classical correlations} Similarly to above, we can characterise the set of classical correlations, $\mathcal{S}^\text{C}_\alpha$, subject to an information bound. In this setting, Alice encodes $x$ into a classical message $m \in[d] $. Bob then provides an output based on his input $y$ and the message $m$. Considering again shared randomness, the resulting correlations take the form
\begin{equation}
p(b|x,y) = \sum_{\lambda}p(\lambda)\sum_{m=1}^d p_\text{A}(m|x,\lambda) p_\text{B}(b|m,y,\lambda).
\end{equation}
In order to characterise correlations of the above form such that $\mathcal{I}_X\leq \alpha$, we proceed as follows. First, 
notice that the dimension $d$ of the message may a priori be unbounded. However, it turns out that, without loss of generality, one can restrict to the case $d=n$. Next, notice that each encoding of the message $p_\text{A}(m|x,\lambda)$ can be taken to be deterministic, i.e.~$m$ is a deterministic function of $x$ and $\lambda$. Finally, to each of these deterministic encodings, we can associate a guessing probability $P_g^{(\lambda)}$. A detailed discussion is given in Appendix.

With these in hand, we notice that the constraint $\mathcal{I}_X\leq \alpha$ is equivalent to $\sum_{\lambda}p(\lambda)P_g^{(\lambda)}\leq 2^{\alpha-H_\text{min}(X)}$, which is linear in $p(\lambda)$. Therefore, the set $\mathcal{S}^\text{C}_\alpha$ forms a convex polytope. The facets of the polytope correspond to linear inequalities 
\begin{equation}\label{witness}
\sum_{x,y,b}r_{xyb} \,p(b|x,y) \leq \beta
\end{equation} 
where $r_{xyb}$ and $\beta$ are real coefficients, which give a complete characterisation of $\mathcal{S}^\text{C}_\alpha$. 

We have explicitly characterised $\mathcal{S}^\text{C}_\alpha$ for scenarios featuring a small number\footnote{Typically, characterising $\mathcal{S}^\text{C}_\alpha$ quickly becomes computationally demanding as we increase the number of inputs and outputs (the number of vertices grows rapidly). While we could solve cases with $n=2,3$ efficiently and the case of $n=4$ preparations within reasonable time, evaluating $n=5$ preparations becomes time-consuming on a standard desktop computer.} of inputs and outputs. We find three types of facet inequalities: (i) positivity conditions, e.g. $p(b|x,y)\geq 0$, (ii) inequalities ensuring the information bound on the observed correlations, e.g. $\sum_x p(b=x|x,y) \leq 2^{\alpha-H_\text{min}(X)}$ (assuming here $n=k$), and (iii) other inequalities. Inequalities (i) and (ii) are in a sense trivial, as they must be satisfied by all physical correlations (when assuming $\mathcal{I}_X\leq \alpha$). On the contrary, inequalities (iii) are non-trivial, and thus capture limits of classical correlations. These inequalities do not necessarily hold for quantum correlations, as we show below. 

Finally, note that the problem of determining whether some observed correlations $p(b|x,y)$ can be obtained classically with $\mathcal{I}_X \leq \alpha$ bits of information is a linear program. One can thus determine the minimal amount of information required to produce $p(b|x,y)$ in a classical protocol.

\section{Quantum advantage} A critical question is whether informationally restricted quantum correlations can outperform their classical counterparts (and thereby provide a quantum advantage). To answer this question, we have considered simple scenarios -- labelled by the number of inputs and outputs, i.e. $(n,l,k)$ -- and characterised their classical polytope $\mathcal{S}^\text{C}_\alpha$. Alice's input is always chosen to be uniformly distributed, i.e. $p_X(x)=1/n$. The simplest scenario where we could find a non-trivial facet inequality is (3,2,2). We conjecture that $n\geq 3$ is necessesary (we have checked that no quantum advantage is possible for $(2,2,2)$ and $(2,2,3)$). 

The scenario $(3,2,2)$ features two non-trivial facets showing a quantum advantage (see Appendix) . Here we focus on one of them:
\begin{equation}\label{ineq}
F_1 \equiv -E_{11}-E_{12}-E_{21}+E_{22}+E_{31}\leq 2^{\alpha+1}-1
\end{equation}
where $E_{xy}= p(0|x,y)-p(1|x,y)$ and $\mathcal{I}_X \leq \alpha\in[0,\log 3]$. Notice that for $\alpha=1$, this inequality is identical to the simplest dimension witness of Ref. \cite{Gallego} for classical bits. 

\begin{figure}
	\centering
	\includegraphics[width=0.95\columnwidth]{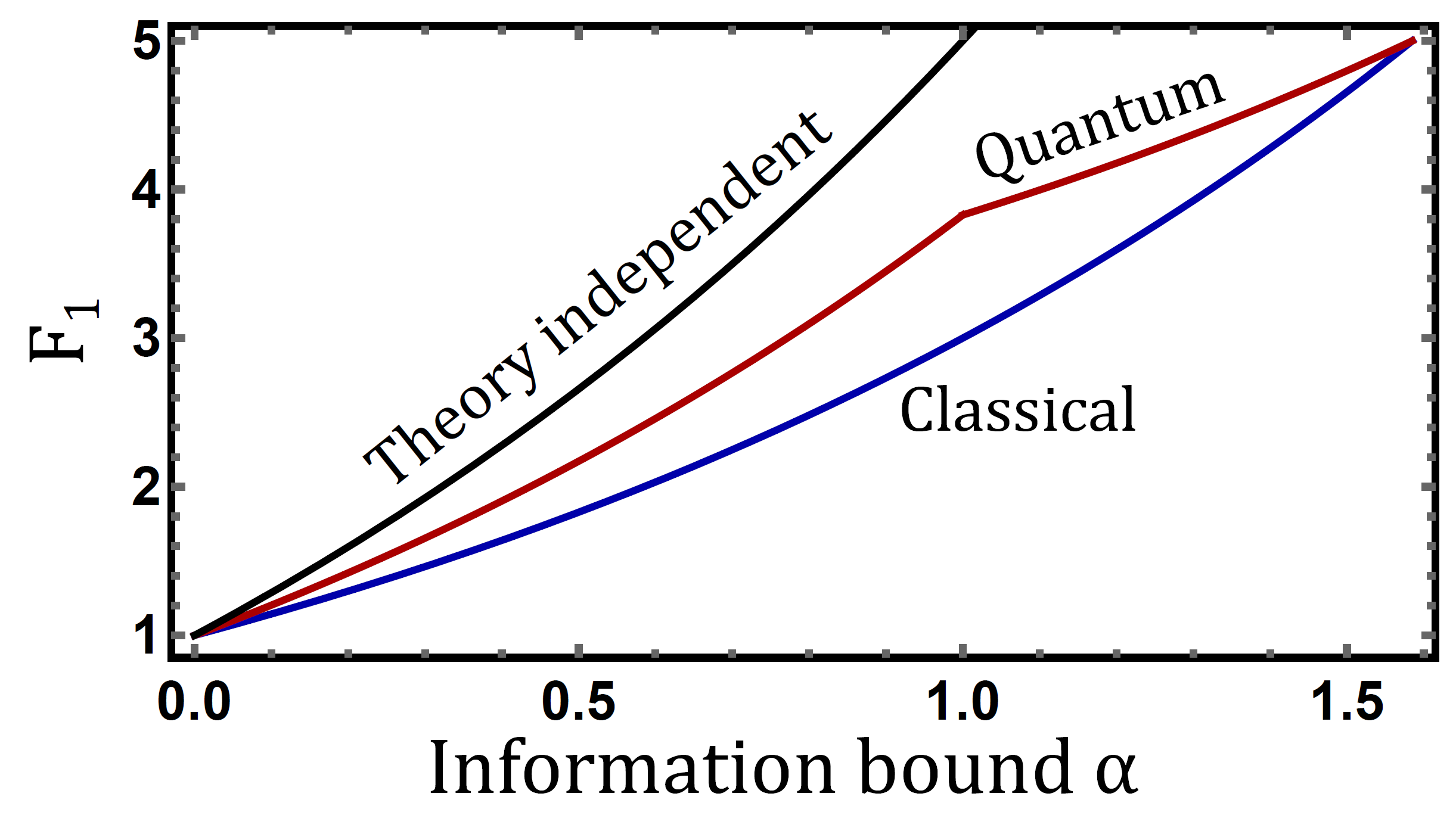}
	\caption{Witness value $F_1$ as a function of the information bound $\mathcal{I}_X \leq \alpha$. Classical correlations necessarily satisfy the inequality \eqref{ineq} (blue curve). Quantum correlations outperform classical ones for $\alpha \in (0,\log3)$; the red curve is obtained by a family of quantum protocols and therefore constitutes a lower bound on the optimal quantum correlations. The black curve represents a lower bound on theory-independent correlations, i.e.~a lower bound on the information needed for the value $F_1$.}\label{FigQ}
\end{figure}

Importantly, the above inequality can be violated in quantum theory whenever\footnote{The extremal cases $\mathcal{I}_X\in\{0,\log3\}$ are trivial since they correspond to no information and relaying $x$ respectively.} $\mathcal{I}_X\in (0,\log3)$, as illustrated in Fig.~\ref{FigQ}. Let Alice and Bob share one bit of randomness ($\lambda\in\{0,1\}$) with distribution $q\equiv p(\lambda=0)$. 
When $\lambda=0$, Alice prepares the qubit ensemble $\mathcal{E}_0=\{\frac{1}{3},\ket{\psi_x}\}$ with $\ket{\psi_1}=\frac{1}{\sqrt{2}}\left(\ket{0}+\ket{1}\right)$, $\ket{\psi_2}=\ket{0}$ and $\ket{\psi_3}=\sin\frac{\pi}{8}\ket{0}-\cos\frac{\pi}{8}\ket{1}$. Bob measures the observables $-\frac{\sigma_x+\sigma_z}{\sqrt{2}}$ and $\frac{\sigma_z-\sigma_x}{\sqrt{2}}$, where $(\sigma_x,\sigma_y,\sigma_z)$ are the Pauli matrices. 
When $\lambda=1$, Alice sends no information and Bob outputs $b=1$ regardless of $y$. This strategy results in the witness value $F_1=1+2\sqrt{2}q$, while the information is $\mathcal{I}_X=\log(1+q)$. Thus, this strategy is relevant in the range $\mathcal{I}_X\in[0,1]$. When $\mathcal{I}_X\in[1,\log(3)]$, we consider another mixed strategy. For $\lambda=0$ we use again the ensemble $\mathcal{E}_0$ and associated measurements, and for $\lambda=1$ a qutrit strategy in which Alice sends $x$ to Bob, thus attaining the maximal value of $F_1=5$. We get $F_1=\left(1+2\sqrt{2}\right)q + 5\left(1-q\right)$ and $\mathcal{I}_X=\log(3-q)$.

An interesting question is to find the optimal value of $F_1$ for any possible quantum strategy with bounded information. This is a non-trivial question as one should consider quantum systems of arbitrarily large Hilbert-space dimension. Based on numerical search, we show in Appendix the existence of slightly better quantum strategies than the above one, but we did not find a simple parameterisation for them.

\section{Device-independent bounds on information} While determining the limits of quantum correlations for limited information is challenging, we can nevertheless infer a general, theory-independent, lower bound on information given observed correlations $p(b|x,y)$. 

The assumption $\mathcal{I}_X \leq \alpha$ implies that, from any of the distributions $\{p(b|x,1),\ldots,p(b|x,l)\}$, one cannot extract more than $\alpha$ bits of information about $x$. Allowing for an arbitrary post-processing of the data (Bob creating a new output $b'$ from $y$ and $b$ ), i.e. $p(b'|y,b)\geq 0$ with $\sum_{b'} p(b'|y,b)=1$ where $b'\in[n]$, we obtain the constraints 
\begin{align}
\forall y : \quad \sum_{x,b}p_X(x)p(b|x,y)p(b'=x|y,b) \leq  2^{\alpha-H_\text{min}(X)}. \nonumber
\end{align}
Determining whether a given correlation $p(b|x,y)$ is compatible with the above constraints can be cast as a linear program. If the program admits no feasible solution, then an information $\mathcal{I}_X>\alpha$ is necessary to reproduce $p(b|x,y)$. Note that, while the above constraints are necessary to ensure that $\mathcal{I}_X \leq \alpha$, they are most likely not sufficient in general. How to derive stronger constraints on information is an interesting open problem.

To illustrate the relevance of these ideas, we have derived a lower bound on $\mathcal{I}_X$ given an observed value of the witness $F_1$. The results are illustrated in Fig.~\ref{FigQ} and demonstrate the possibility of certifying a device-independent lower bound on the information. Note that the bound applies to quantum correlations, and more generally to any operational theory.

\section{Information versus dimension} Another relevant question is to compare quantum correlations with bounded information to those achievable with bounded dimension. Such comparison makes sense when $\mathcal{I}_X \leq  \log{d}$, where $d$ is the Hilbert-space dimension of the quantum systems. Clearly, any correlation achieved via $d$-dimensional systems (qudits) requires at most $\mathcal{I}_X = \log{d}$, as any ensemble of qudits carries no more than $\log{d}$ bits of information \cite{Holevo}. However, it turns out that there are quantum correlations not achievable via qudits that can nevertheless be obtained with information $\mathcal{I}_X = \log{d}$.

Specifically, we consider the case $d=2$ and exhibit quantum correlations achievable with $\mathcal{I}_X =1$ that cannot be obtained from qubits. Consider a Random Access Code \cite{Ambainis, RACthesis, TH15} in which Alice receives a uniformly random four-bit input $x=(x_1,x_2,x_3,x_4)\in[2]^4$. Bob has settings $y\in[4]$, and returns a binary output $b$ with which he aims to guess $x_y$. The score is
\begin{equation}\label{RAC}
F_{\text{RAC}}=\frac{1}{64}\sum_{x,y}p(b=x_y|x,y).
\end{equation}
Qubit strategies must satisfy $F_{\text{RAC}} < 3/4$; this follows from the impossibility of having four mutually unbiased bases for qubits \cite{RACthesis, TK18}. Moreover, numerical optimisation strongly suggests that 
$F_{\text{RAC}}\leq 0.741$ for qubits \cite{RACthesis}.

It is nevertheless possible to obtain the score $F_{\text{RAC}} = 3/4$ using quantum ensembles with $\mathcal{I}_X =1$. The strategy employs 16 four-dimensional quantum states of the form
\begin{multline}\label{ens}
\rho_x=\frac{1}{8}\bigg(2\openone\otimes \openone-(-1)^{x_4}\openone\otimes\sigma_y-(-1)^{x_1}\sigma_x\otimes\sigma_x\\
 -(-1)^{x_2}\sigma_y\otimes\sigma_x-(-1)^{x_3}\sigma_z\otimes\sigma_x\bigg),
\end{multline}
and Bob measures the observables $B_1=\sigma_x\otimes \sigma_x$, $B_2=\sigma_y\otimes \sigma_x$, $B_3=\sigma_z\otimes \sigma_x$ and $B_4=\openone\otimes \sigma_y$.
Note that, despite being four-dimensional, these states are noisy (with purity $\Tr\left(\rho_x^2\right)=1/2$ $\forall x$) and carry only one bit of information. Since all states have the same spectrum, $(1/2,1/2,0,0)$, this can be checked analytically as follows. For any quantum ensemble, the information is upper bounded by 
\begin{equation}\label{bound}
I_X\leq\log\left(d\right)+\log \left(\frac{\max_x p_X(x)\lambda_\text{max}(\rho_x)}{\max_x p_X(x)}\right),
\end{equation}
where $\lambda_\text{max}(\rho_x)$ is the largest eigenvalue of $\rho_x$, and $d$ the Hilbert-space dimension. The bound is obtained from using the relation $\Tr\left[\rho_xN_x\right]\leq \lambda_\text{max}(\rho_x)\Tr\left[N_x\right]$ in Eq.~\eqref{guessing} and then $\sum_x N_x=\openone_d$. The bound Eq.~\eqref{bound} is tight when (i) for each $x$, $\rho_x$ only has one non-zero eigenvalue (with possible multiplicity) and (ii) $p_X(x)\lambda_\text{max}(\rho_x)$ is constant in $x$. The ensemble in Eq.~\eqref{ens} satisfies this criteria.

An interesting question is whether larger separation  is possible. That is, how much stronger can quantum correlations with $\mathcal{I}_X = \log{d}$ bits of information become compared to quantum correlations using $d$-dimensional quantum systems. In Appendix, we show that, in a scenario without shared randomness, this advantage can  be made arbitrarily large. Specifically, we construct quantum correlations achievable with $\mathcal{I}_X = 1$ bit of information, that can only be reproduced using an arbitrary large Hilbert-space dimension.

\section{Quantum communication versus entanglement-assisted classical communication} Informationally restricted quantum systems also have interesting implications when comparing quantum resources in different communication scenarios. On the one hand, Alice may send an amount of quantum communication to Bob (as in Fig. 1). On the other hand, Alice and Bob may share unlimited entanglement and Alice communicates the same amount classically. These two approaches are generally not equivalent. In fact, for dimensionally restricted classical and quantum messages, there is no strict hierarchy. In some cases, quantum communication outperforms entanglement-assisted classical communication \cite{Tavakoli1, SpatialSequential, HS17} and vice versa in others \cite{PZ10, HS17, TZ17}. Given this seemingly complicated picture, no generally valid criterion is known for determining which quantum resource is more efficient for a given communication task. Interestingly, we show that every correlation obtained via entanglement-assisted classical communication of a $d$-dimensional message can also be obtained via quantum communication carrying at most $\log d$ bits of information. That is, in this setting, quantum communication is the stronger resource.

Consider a scenario with classical communication, where Alice and Bob can use a pre-shared entangled state $\rho_{\text{AB}}$. Upon receiving input $x$, Alice performs a measurement $\{A_{a|x}\}$ with outcome $a$ on her half of $\rho_{\text{AB}}$, which projects Bob's system onto the state $\sigma_{a|x} = \Tr_\text{A}([A_{a|x} \otimes \openone_\text{B}]\rho_{\text{AB}})/p(a|x)$, where $p(a|x) = \Tr([A_{a|x} \otimes \openone_\text{B}]\rho_{\text{AB}})$. Alice then sends a classical message to Bob; which we represent as a collection of $d$-dimensional quantum states $\mu_{a|x}$ diagonal in the same basis. Thus, Bob holds the classical-quantum state $\mu_{a|x} \otimes \sigma_{a|x}$, on which he can perform some measurements in order to establish correlations $p(b|x,y)$. The information cost of this protocol originates only from the classical message, as the entanglement is pre-shared.

Now, we construct a quantum communication protocol to simulate the above correlations using at most $\log d$ bits of information. Upon receiving $x$, Alice samples from $p(a|x)$, and sends to Bob the classical-quantum state $\mu_{a|x} \otimes \sigma_{a|x}$. Evidently, Bob can now reproduce the same correlations $p(b|x,y)$. The key point is now to show that this protocol does not require more information than above. The ensemble (averaged over $a$) can be written $\mathcal{E}_\text{QC}=\{p_X(x),\tau_{x}\}$ where $\tau_x=\sum_a p(a|x) \mu_{a|x} \otimes \sigma_{a|x}$. The corresponding guessing probability is
\begin{align}\label{exp}
P_g^{\text{QC}}=\max_{\{N_z\}} \sum_{a,x} p_X(x) p(a\lvert x)  \Tr \left( \mu_{a|x} \otimes \sigma_{a|x} N_x\right) 
\end{align}
where the POVM $\{N_z\}$ acts jointly on the classical message space and on the quantum state space. We can place the following upper bound on the guessing probability
\begin{equation}
P_g^\text{QC}\leq \max_{\{N_z\}} \sum_{x} p_X(x) \Tr \left(  \left(\sum_a p(a\lvert x) \sigma_{a|x}\right)  N_x^\text{B}\right),
\end{equation}
where we have used that $\Tr ( \mu_{a|x} \otimes \sigma_{a|x} N_x)  \leq \Tr ( \sigma_{a|x}  N_x^\text{B}) $, where $N_x^\text{B} $ is the partial 
trace of $N_x$ over the first system (the classical message space). Importantly,  since for every $x$ the ensemble $\{p(a|x),\sigma_{a|x}\}$ is remotely prepared by Alice on Bob's side, it follows that 
\begin{multline}
\sum_a p(a|x)\sigma_{a|x}=\sum_a \Tr_\text{A}\left(A_{a|x}\otimes \openone_\text{B} \rho_\text{AB}\right)\\
=\Tr_\text{A}\left(\rho_\text{AB}\right)=\rho_\text{B}.
\end{multline}
Therefore, the guessing probability obeys
\begin{align}
& P_g^\text{QC}\leq \max_{\{N_z\}} \sum_{x} p_X(x) \Tr \left( N_x^\text{B}\rho_\text{B}\right)\\
& \leq \left(\max_xp_X(x)\right) \max_{\{N_z\}}\Tr\left(\sum_x N_x^\text{B}\rho_\text{B}\right). 
\end{align}
Finally, we use the completeness relation of POVMs to obtain
\begin{equation}
\sum_x N_x^\text{B}=\sum_x \Tr_\text{1}\left(N_x\right)=\Tr_\text{1}\left(\openone_d\otimes \openone\right)=d\openone,
\end{equation}
where we have used that the identity operator on the classical message space is $d$-dimensional. Thus, we conclude that
\begin{align}
& P_g^\text{QC}\leq  d \max_xp_X(x). 
\end{align}
Consequently, the information is bounded by
\begin{align}\nonumber
& \mathcal{I}_X=-\log\left(\max_xp_X(x)\right)+\log\left(P_g^\text{QC}\right)\\
& \leq -\log\left(\max_xp_X(x)\right)+\log\left(d \max_xp_X(x)\right)=\log d.
\end{align}
This concludes the proof: quantum communication of $\log d$ bits of information is a stronger resource than classical communication of a $d$-dimensional message assisted by any amount of entanglement. 

Finally, we also note that this proof remains valid also if Alice uses her classical outcome $a$ and her input $x$ to encode a quantum $d$-dimensional message $\mu_{a|x}$. This is, however, not the most general quantum operation that may be considered.

\section{Conclusions}  We have investigated correlations in prepare-and-measure scenarios under the assumption of an upper bound on the information. We have shown how to fully characterise correlations in the case of classical systems and proved a quantum advantage. Moreover, we showed that stronger quantum correlations can be obtained when bounding the information rather than the dimension, and devised device-independent tests of information. 
	
An outstanding open question is to characterise quantum correlations when the transmitted information is bounded. Is it sufficient to consider quantum ensembles of finite dimension, as in the classical case? Or are there correlations that require infinite-dimensional quantum systems? Another point is to understand how much stronger quantum correlations can be compared to classical ones. For the case where shared randomness is not allowed, we could show a diverging advantage. Is it also the case in a scenario including shared randomness? In addition, it would be interesting to consider informationally restricted correlations based on other information measures than the one we consider here; for instance based on Shannon entropies. Another possible direction is to explore connections between our approach and other scenarios in information theory, for instance the (quantum) information bottleneck function \cite{Tishby,Datta}. Furthermore, it would also be relevant to explore the role of informationally restricted correlations with respect to the line of research focused on operational contextuality \cite{Spekkens} in which one considers prepare-and-measure scenarios featuring an assumption of oblivious communication (see e.g.~\cite{POM, TavakoliContext}).

Finally, we briefly discuss the prospects of using our approach in experiments, notably towards possible applications in semi-device-independent (SDI) quantum information processing. In this area, protocols so far were mostly based on a dimension assumption, see e.g. \cite{Pawlowski, LY11, WP, Lunghi, TK18}, which is usually justified from the physics of the experiment. For instance, a setup where the relevant degree of freedom is the polarization of a single photon motivates the assumption that the prepared states can be described as qubits. In practice, however, single-photon sources feature imperfections which result in unavoidable multi-photon emissions, which clearly no longer satisfy the qubit assumption. Taking these into account is typically cumbersome and inefficient (see for instance \cite{Lunghi}). In comparison, the information approach might be much better adapted here. From a physical model of the source, the rate of multi-photon events can be estimated. For instance, a weak laser source will exhibit Poisson statistics. For each photon number the carried information can be estimated, which in turn results in an overall bound on the carried information. In this way, one could continuously tune the information bound, taking into account the relevant degrees of freedom and photon statistics. Bounding the information rather than the dimension may therefore represent a more natural assumption, better motivated by the physics of the source. It would be interesting to explore these ideas in practice, as well as to understand the relation between the information approach and other SDI approaches recently developed, based on bounding the energy \cite{vanHimbeeck}, the overlap \cite{Brask, Charles} or the entropy \cite{Chaves} of the quantum communication.

\begin{acknowledgements}
 We thank Jean-Daniel Bancal, Joseph Renes, Renato Renner, Joseph Bowles, Alastair Abbott, Francesco Buscemi, Michele Dall'Arno, Stefano Pironio and Erik Woodhead for discussions. This work was supported by the Swiss National Science Foundation (Starting grant DIAQ, NCCR-QSIT) and the Independent Research Fund Denmark.
\end{acknowledgements}

\appendix

\section{Characterisation of classical correlations}\label{AppClassical}

We describe a classical scheme, starting with deterministic strategies. Alice uses an encoding function $E:[n]\rightarrow [d]$ to associate her input to a $d$-valued message $m=E(x)$ and sends it to Bob. No limitation on $d$ is assumed. Bob uses a decoding function $D:[d]\times [l]\rightarrow [k]$ to map the pair $(m,y)$ into an $k$-valued output $b=D(m,y)$. Since there are $Z_\text{A}=d^{n}$ ($Z_\text{B}=k^{dl}$) possible encoding (decoding) functions, the number of deterministic strategies is $Z=Z_\text{A}Z_\text{B}$. We index them by $(E_{\lambda_\text{A}},D_{\lambda_\text{B}})$ for $\lambda_\text{A}\in[Z_\text{A}]$ and $\lambda_\text{B}\in [Z_\text{B}]$ respectively. Via the shared randomness $\lambda=(\lambda_\text{A},\lambda_\text{B})$, classical correlations are written
\begin{equation}\label{cprob}
p^\text{C}(b|x,y)=\sum_{\lambda}p(\lambda)\sum_{m=1}^d \delta_{m,E_{\lambda_\text{A}}(x)}\delta_{b,D_{\lambda_\text{B}}(m,y)}.
\end{equation}
We now characterise $p^\text{C}(b|x,y)$ when $\mathcal{I}_X\leq \alpha$ for some real $\alpha\geq0$. To this end, we need to eliminate the dimension $d$. Below, in section~\ref{dim} we show that without loss of generality one can choose $d=n$ (i.e.~the dimension equal to the number of inputs for Alice). We will use this fact to characterise the polytope of classical correlations and leave the proof for the end of this section.

\subsection{The classical polytope}
We use that classical messages of dimension $d=n$ are sufficient. Therefore, we can denote all encoding functions and decoding functions $(E_{\lambda_\text{A}},D_{\lambda_\text{B}})$ where the index $\lambda=(\lambda_\text{A},\lambda_\text{B})$ acts as a shared random variable (whose cardinality is now finite) allowing the coordination of deterministic encoding and decoding strategies. For a fixed deterministic strategy, we obtain a distribution $p'_\lambda(b|x,y)$. This distribution is a vertex of the polytope $\mathbb{P}$ which is the space of all probabilities $p(b|x,y)$. However, many deterministic strategies give rise to the same vertex in the probability space. Therefore, we write $\{p_\gamma(b|x,y)\}_\gamma$ for the unique elements in the set $\{p_\lambda'(b|x,y)\}_\lambda$. We define 
\begin{equation}
E_\gamma=\{\lambda=(\lambda_\text{A},\lambda_\text{B})|p_{\gamma}(b|x,y)=p_{\lambda}'(b|x,y)\},
\end{equation}
where $\{p_\gamma(b|x,y)\}$ is the list of vertices of $\mathbb{P}$ (without duplicates). In other words, $E_\gamma$ is the set of all deterministic strategies that generate the vertex $p_{\gamma}(b|x,y)$. 

To each vertex of $\mathbb{P}$ we associate the smallest amount of  information needed to generate it (for simplicity, we work with the guessing probability). That is,  
\begin{equation}
P_g^{(\gamma)}=\min_{\lambda\in E_\gamma}P_g^{(\lambda_\text{A})}
\end{equation}
where the guessing probability of the deterministic strategy is given by
\begin{equation}\label{cguess}
P_g^{(\lambda_\text{A})}=\max_{\mu} \sum_{x}p_X(x)\sum_{m=1}^{d}\delta_{m,E_{\lambda_\text{A}}(x)}\delta_{x,\tilde{D}_{\mu}(m)},
\end{equation} 
where the maximisation is over all the deterministic decoding strategies $\tilde{D}:[d]\rightarrow [n]$ (of which there are $n^d$).

We now impose the information restriction,  $I_X\leq \alpha$. This can be formulated as a linear constraint in the shared randomness. The characterisation of the set of information restricted classical correlations reads  
\begin{align}
& p(b|x,y)=\sum_{\gamma}p(\gamma)p_\gamma(b|x,y)\\
& \sum_{\gamma}p(\gamma)P_g^{(\gamma)}\leq 2^{\alpha-H_\text{min}(X)}\\
& \sum_{\lambda} p(\gamma)=1 \\
& p(\gamma)\geq 0.
\end{align}
This defines a convex polytope. Its facets can be obtained using standard polytope software. We label this polytope $\mathbb{P}_\alpha$ and note that it is contained inside $\mathbb{P}$.

\begin{figure}[H]
	\centering
	\includegraphics[width=70mm,scale=0.03]{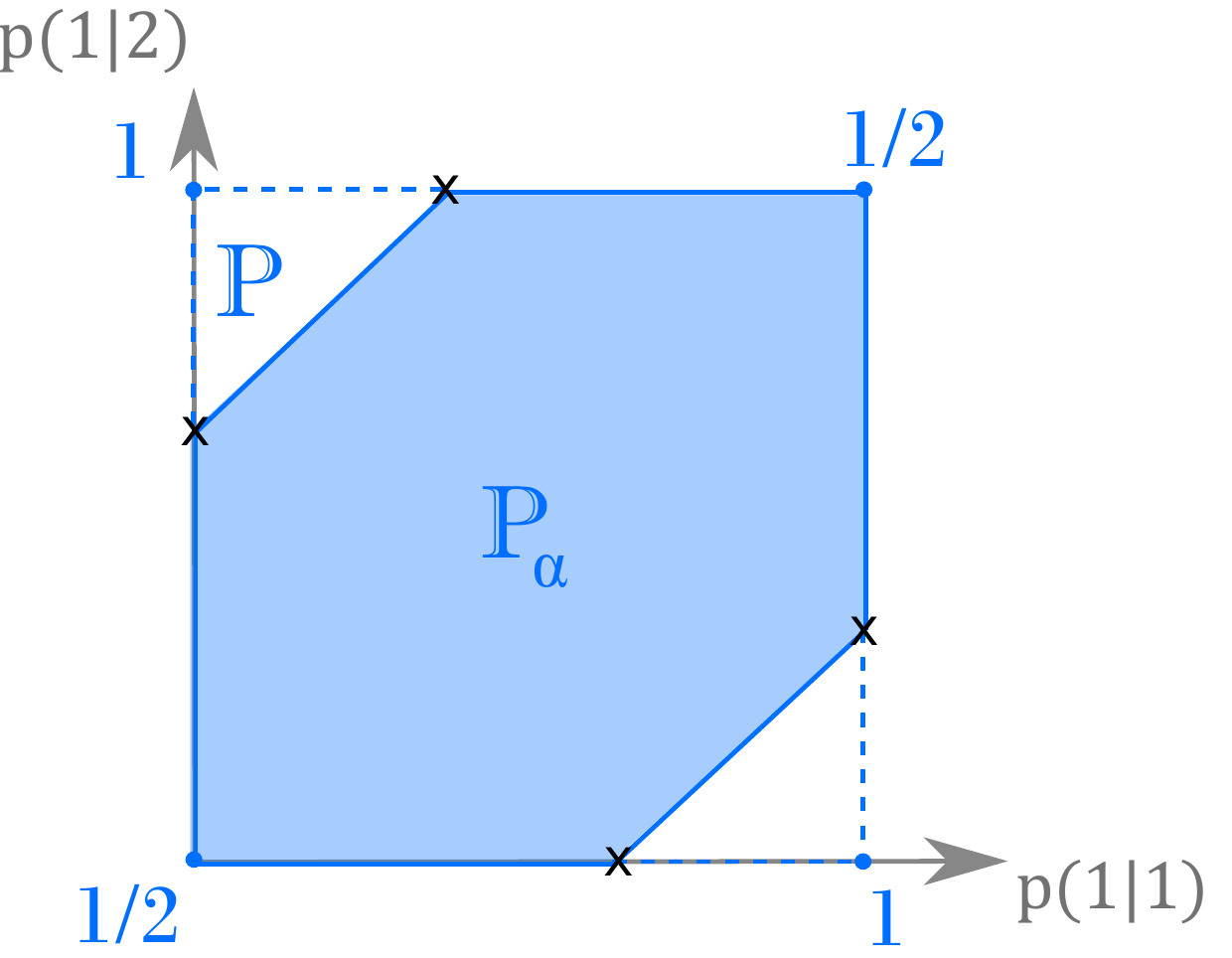}
	\caption{The classical set of correlations for a scenario with two preparations and one binary outcome measurement $(n,l,k)=(2,1,2)$. The polytope $\mathbb{P}$ has four vertices, each corresponding to a guessing probability of either one or one half (written in blue). The facets are lines. Therefore there is only one pair of vertices per facet, for each of which we inscribe a new vertex (represented by a tick) as imposed by limiting the guessing probability. Thus, the blue region is the polytope $\mathbb{P}_\alpha$.}
	\label{CutPolytope}
\end{figure}

As an illustration of how the polytope $\mathbb{P}_\alpha$ may look, we have displayed in Fig.~\ref{CutPolytope} a schematic of the polytope in the simplest case of Alice having two inputs and Bob performing a single binary outcome measurement ($n=k=2$, $l=1$), for which the polytopes $\mathbb{P}$ and $\mathbb{P}_\alpha$ are polygons.

\subsection{Optimal classical correlations via linear programming}
Since the set of classical correlations forms a convex polytope for $\mathcal{I}_X\leq \alpha$, one can determine whether a given $p(b|x,y)$ belongs to said polytope via a linear program. This allows one to determine whether $p(b|x,y)$ is classically realisable with information no more than $\alpha$.

Moreover, given any linear functional of probabilities, 
\begin{equation}\label{F}
F=\sum_{x,y,b}r_{xyb} \,p(b|x,y),
\end{equation}
one can determine the exact classical bound through the evaluation of the linear program
\begin{align}\label{LP}\nonumber
& F^\text{C}=\max_{p(\lambda)} F[p(b|x,y)] \\\nonumber
& \text{such that } \sum_{\lambda}p(\lambda)P_g^{(\lambda_\text{A})}\leq 2^{\alpha-H_\text{min}(X)},\\
& \sum_{\lambda}p(\lambda)=1, \quad \text{and} \quad  p(\lambda)\geq 0.
\end{align}
This allows to obtain witnesses for classical correlations.

\subsection{Dimension $n$ is sufficient for classical messages}\label{dim}

Here, we show that the optimal classical correlations, for any correlation witness constrained by bounded guessing probability (or equivalently, bounded information) with shared randomness, is obtained with a message dimension not larger than the cardinality of the input of Alice, i.e.~$d=n$.

Any classical strategy can be decomposed as a mixture of deterministic strategies, as given by Eq.~\eqref{cprob}. For a fixed value of the shared variable $\lambda$, the encoding strategy $E_{\lambda_\text{A}}$ is fixed. Since $x$ can take at most $n$ different values, there is then at most $n$ different values of $E_{\lambda_\text{A}}(x)$. Thus, for a fixed $\lambda$, at most $n$ message symbols are used. Whether there is any advantage in using message dimensions $d>n$ thus becomes a question of whether there is any advantage in using different sets of message symbols for different $\lambda$. 

We first show that any value of the maximum in Eq.~\eqref{LP} obtained with different sets of message symbols for different $\lambda$ can also be achieved using the same set of $n$ symbols for all $\lambda$. This can be seen from Eq.~\eqref{cprob}. For each  value of $\lambda$, the factor $\delta_{m,E_{\lambda_\text{A}}(x)}$ is nonzero for at most $n$ different values of $m$. The decoding function $D_{\lambda_\text{B}}(m)$ hence needs to be defined only on these values. If any of these values lie outside $[n] = \{1,\ldots,n\}$ then there must be corresponding values in $[n]$ which are not used. We can then redefine $E_{\lambda_\text{A}}$ and $D_{\lambda_\text{B}}$ to use these values instead. 

Specifically, for some fixed $\lambda$, say that $E_{\lambda_\text{A}}(x_0) = \nu \notin [n]$ for some $x_0$. Then there exists $\nu'\in[n]$ such that $E_{\lambda_\text{A}}(x) \neq \nu'$ for all $x$. We then define
\begin{align}
E_{\lambda_\text{A}}'(x) & = \begin{cases}
\nu' & \text{if} \,\, x=x_0 , \\
E_{\lambda_\text{A}}(x) &\text{otherwise} ,
\end{cases} \\
D_{\lambda_\text{B}}'(m) & = \begin{cases}
D_{\lambda_\text{B}}(\nu) & \text{if} \,\, m=\nu' , \\
D_{\lambda_\text{B}}(m) &\text{otherwise} .
\end{cases}
\end{align}
Substituting $E_{\lambda_\text{A}} \rightarrow E_{\lambda_\text{A}}'$ and $D_{\lambda_\text{B}} \rightarrow D_{\lambda_\text{B}}'$ in \eqref{cprob} leaves the probabilities $p(b|x,y)$ unchanged. Repeating this process, the message symbols can be chosen in $[n]$ for every $\lambda$, without changing the probabilities and hence a distribution achieving the optimum in Eq.~\eqref{LP} remains optimal.

The only remaining question is now, whether this remapping to a strategy using the same $n$ symbols for all $\lambda$ can lead to violation of the information constraint. From \eqref{cguess}, we can see that this is not the case. Let $\tilde{D}_{\mu^*}$ be the optimal decoding function which achieves the maximum on the right-hand side of \eqref{cguess}, for some fixed $\lambda$. When $E_{\lambda_\text{A}}$ is replaced by $E_{\lambda_\text{A}}'$ as above, the maximum remains unchanged and is achieved by
\begin{align}
\tilde{D}_{\mu^*}'(m) & = \begin{cases}
\tilde{D}_{\mu^*}(\nu) & \text{if} \,\, m=\nu' , \\
\tilde{D}_{\mu^*}(m) &\text{otherwise} .
\end{cases}
\end{align}
Thus, following the recipe above, we can replace all the encoding and decoding functions $E_{\lambda_\text{A}}: [n] \rightarrow [d]$, $D_{\lambda_\text{B}} : [d] \rightarrow [n]$, and $\tilde{D}_{\mu} : [d] \rightarrow [n]$ by other functions $E_{\lambda_\text{A}}: [n] \rightarrow [n]$,  $D_{\lambda_\text{B}} : [n] \rightarrow [n]$, and $\tilde{D}_{\mu} : [n] \rightarrow [n]$ without changing the probabilities $p(b|x,y)$ or the guessing probabilities $P_g^{(\lambda_\text{A})}$. It follows that the optimum of Eq.~\eqref{LP} can always be attained using a message dimension of at most $n$.

\section{Case study for $(n,l,k)=(3,2,2)$}\label{AppCase}
We have obtained the facets of the polytope for several simple scenarios. The simplest scenario in which we have found non-trivial facets is $(n,l,k)=(3,2,2)$. One can consider different values for the information bound $\mathcal{I}_X\leq \alpha$. We have considered different values of $\alpha$ for each of which we have found two non-trivial inequalities (i.e.~they are not positivity nor the information restriction). More precisely, we considered eleven evenly spaced values of the guessing probability in the range $(1/3,1)$. The facets are 
\begin{align}
& F_1=\sum_{x,y}t_{x,y}^1E(x,y)\leq 6P_g-1 \label{ineqB2}\\
& F_2=\sum_{x,y}t_{x,y}^2E(x,y)\leq 12P_g-4. \label{ineqB3}
\end{align}
where $t_{x,y}^1=\{[-1,-1],[-1,1],[1,0]\}$ and $t_{x,y}^2=\{[-1,-1],[-1,1],[2,0]\}$. Note that for convenience, we have expressed the upper bounds in terms of the guessing probability instead of the information. Both inequalities can be violated in quantum theory. For the first inequality, a violation valid for any non-trivial information was presented in the main text using a quantum strategy with one bit of shared randomness. Notably, said strategy also violates the second inequality but not in the entire range $\mathcal{I}_X\in(0,\log3)$. 

\begin{figure}
	\centering
	\includegraphics[width=\columnwidth]{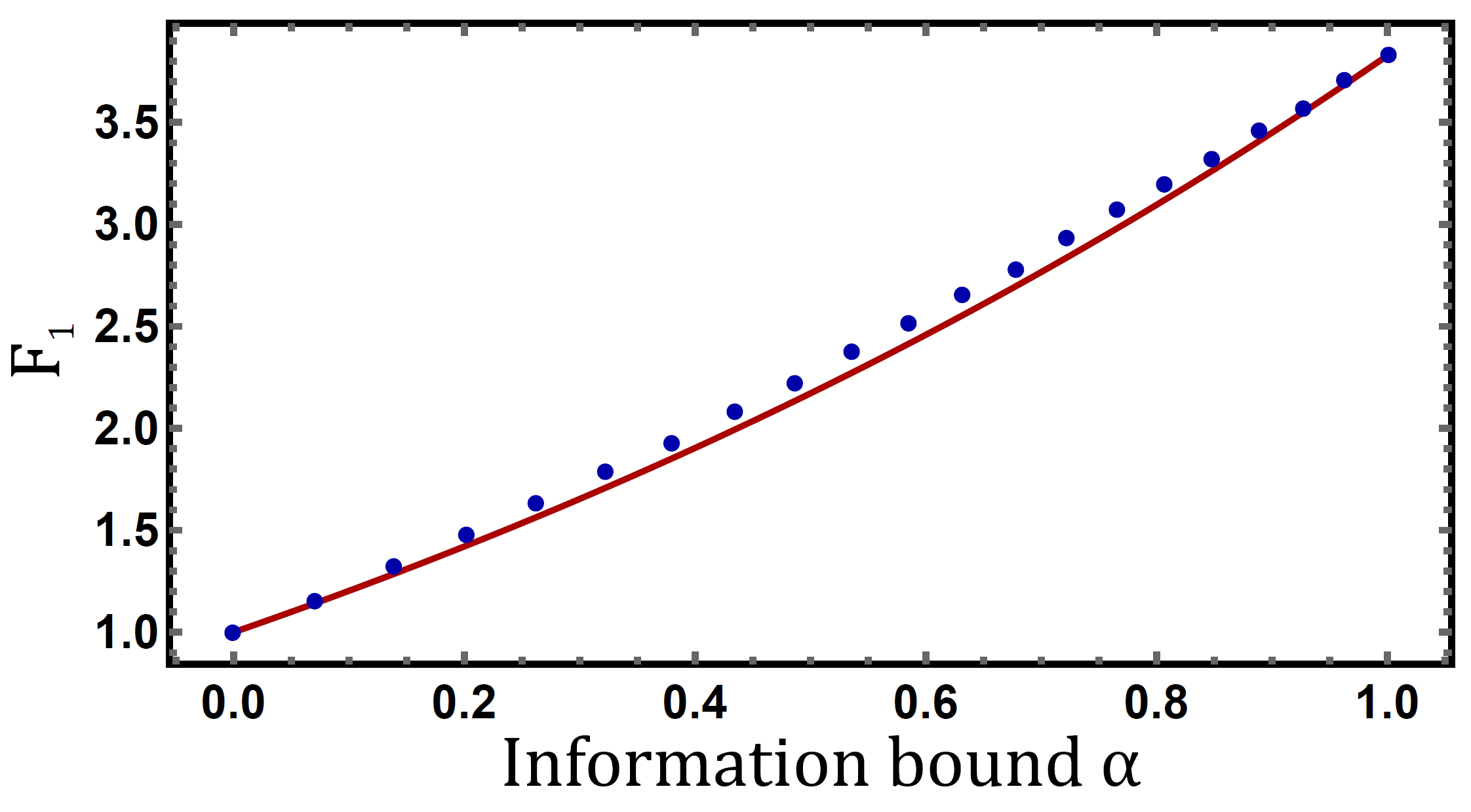}
	\caption{Witness value $F_1$ as a function of the information $\mathcal{I}_X\leq \alpha$. The quantum strategy from the main text is displayed (red curve) and the numerically obtained quantum violations based on qutrits are displayed in blue. In the range $\alpha\in(0,1)$ these improve on the first quantum strategy. Notably, numerics showed that an improvement on the red curve is possible already with qubit preparations.}\label{FigQ2}
\end{figure}

Moreover, we have numerically explored whether larger violations of the first inequality are possible. We considered the case in which Alice prepares general qutrit states and found it to be advantageous. We have employed a brute-force numerical search using the function ``fmincon`` in MATLAB. We employ an effective Lagrange multiplier $\lambda$ and seek to maximise the function
\begin{equation}
\tilde{F}_1=F_1-\lambda |\mathcal{I}_X-\alpha|,
\end{equation}
for a given information bound $\alpha$. We have chosen $\lambda=100$. In every step, we evaluate the information $\mathcal{I}_X$ in the three preparations via a semidefinite program. Then, we evaluate the largest possible value of $F_1$ for the given preparations, which thanks to the binary outcomes can be cast as an eigenvalue problem. We then  ask MATLAB to maximise $\tilde{F}_1$. In Fig~\ref{FigQ2} the results are compared to those of the strategy in the main text. In the range $\mathcal{I}_X\in(0,1)$ we find an improvement, but not in the range $\mathcal{I}_X\in[1,\log3]$. However, we have not found a simple parameterisation of these quantum strategies. Also, it could be possible that even better results can be obtained with higher-dimensional preparations.

\section{Arbitrary large advantage over dimension-bounded quantum ensembles without shared randomness}\label{AppNoSR}
In the main text, we showed that one bit of communication is not always optimally encoded in a qubit ensemble but sometimes in an ensemble of higher-dimensional quantum systems. Here, we show that such advantages over dimension-bounded systems can become more significant in scenarios without shared randomness.

Consider the following variant of a quantum Random Access Code (without shared randomness). Alice has a uniformly random variable $X\in[2^n]$ with values  $x=x_1\ldots x_n\in[2]^n$. She sends $m$ bits of information to Bob, who has a random variable $Y\in[n]$ with values $y$ from which he produces an outcome $b\in[2]$. The aim is to maximise the \textit{worst-case} success probability of finding $b=x_y$, i.e., 
\begin{equation}
\mathcal{A}_{n}^m=\min_{x,y} p(b=x_y\lvert x,y).
\end{equation} 

Let us first choose $m=1$. It is known that with two-valued classical messages or with two-dimensional quantum systems, it is impossible to achieve a better result than that obtained with random guessing, i.e. $\mathcal{A}_n^1=1/2$, when $n>3$ \cite{Hayashi}. In contrast, for $n=2$ and $n=3$, qubits hold an advantage over classical two-valued messages. The reason is that for $n=4$ (and analogously for $n>4$) it is impossible to cut the Bloch sphere into $2^4=16$ symmetric parts with four planes passing through the origin. By a similar argument using the generalised higher-dimensional Bloch sphere, it has been shown \cite{Hayashi} that for general integers $m\geq 1$, sending $m$ classical two-valued messages or sending $m$ qubits ($2^m$-dimensional quantum systems) cannot achieve a better result than $\mathcal{A}=1/2$ when $n$ is choosen as at least $2^{2m}$. 

We compare this with sending a general quantum ensemble of limited information. Again, we first choose $m=1$ and $n=4$. Using the ensemble and measurements specified in the main text for four-bit Random Access Code (average success probability variant), one immediately finds that $\forall x,y: p(b=x_y\lvert x,y)=3/4$, and therefore that $\mathcal{A}_4^1=3/4$. Thus, the ensemble of mixed four-dimensional systems provides an advantage over two-valued classical messages when qubit ensembles fail to provide any better-than-classical result. 

Refs.~\cite{Index, SpatialSequential} derived Bell inequalities for Random Access Codes. Using the results of Ref.~\cite{Index}, Alice and Bob can share an entangled state of local dimension  $D=2^{\lfloor \frac{n}{2}\rfloor}$ and use their inputs as settings for testing the Bell inequalities of \cite{Index, SpatialSequential}. Then, if Alice communicates her binary outcome to Bob, he can satisfy the relation $b=x_y$ with probability
\begin{align}\label{ress}
& \forall x,y: &p(b=x_y\lvert x,y)=\frac{1}{2}+\frac{1}{2\sqrt{n}}.
\end{align}
In the main text we showed that any correlations achievable by means of entanglement-assisted classical communication also is achievable by means of quantum communication without sending more information (and without the need of share randomness). Therefore, we can obtain the correlations  \eqref{ress} using the quantum communication model discussed in the main text. Consequently, using only a single bit of quantum information (encoded in a general ensemble), we can achieve
\begin{align}
\mathcal{A}_n^1=\frac{1}{2}+\frac{1}{2\sqrt{n}}.
\end{align}
Note that this is strictly greater than $1/2$ for all $n\geq 2$. Therefore, if we choose $n\geq 2^{2m}$ but use only a single bit of information, we outperform the best possible quantum protocols in which the allowed $m$ bits are encoded in $2^m$-dimensional quantum systems. Thus, the advantage is unbounded in the sense that a fixed amount (one bit) of general quantum information holds an advantage over the $m$ bits carried by $m$ qubits, for any (potentially) arbitrarily large choice of $m$.

%
%
%
%
%
%
%
%
%
%
%
%
%

\end{document}